**Astronomia di Posizione per Muoni,** *Algoritmi per foglio elettronico*

Costantino Sigismondi

APRA, Roma (2008) ISBN 9788889174852

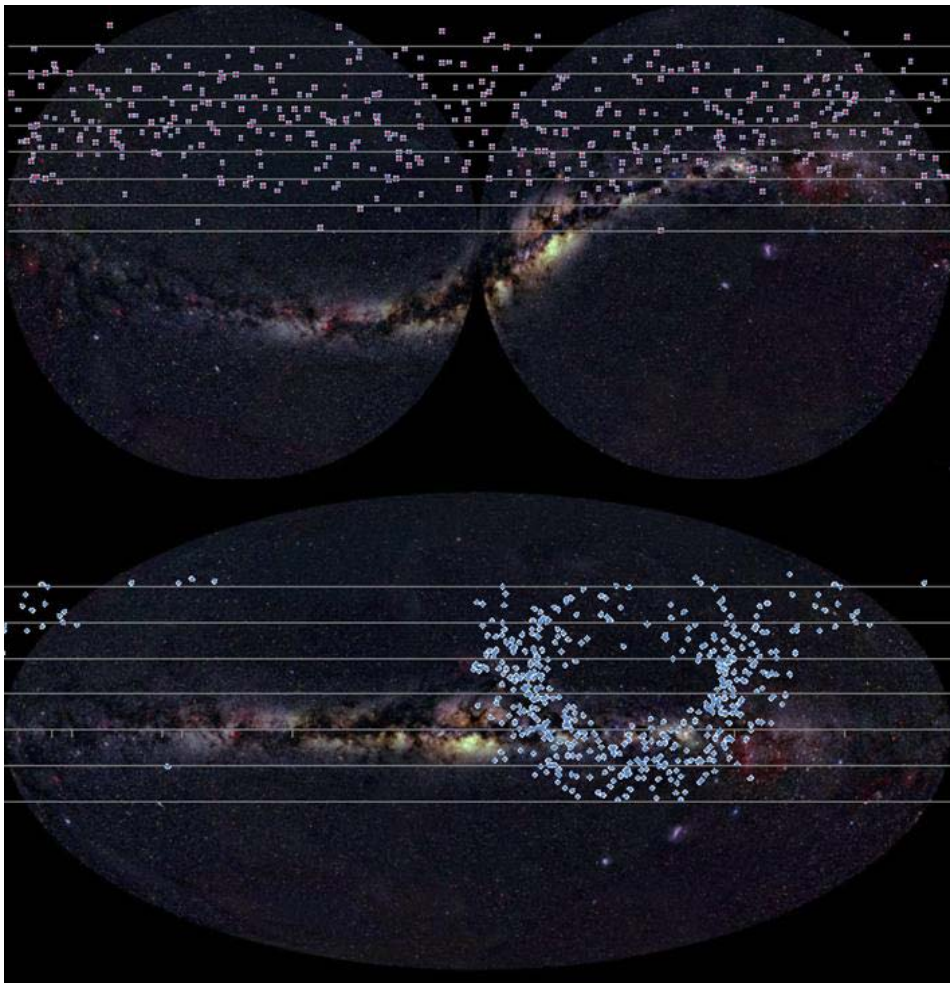

**Positional Astronomy for Muons,** *Algorithms for electronic spreadsheet*

**Abstract :**

The muons of cosmic rays air showers in the Extreme Energy Events (EEE) project are detected with three Multi-gap Resistive Plate Chambers (MRPC) with good tracking capability. These muon telescopes are located in high schools spread all over Italy. The detection of extensive air showers is made by means of time coincidences between two distant telescopes. [1] The vectorial components of the incoming directions of the muons are known, as well as the UTC time of their arrival on the detectors. The method to calculate the celestial (equatorial and galactic) coordinates of the incoming direction of the muons is here presented. This procedure allows recovering galactic or extragalactic sources of the extreme energetic cosmic rays which produce such extensive air showers. A worksheet file (muoni.xls or EEEtest.xls) contains a simulator, to produce data in the same format. This introductory method to positional astronomy for muons, useful also for neutrinos, is presented through explained formulae and an interactive worksheet, tailored for the data format of EEE (http://www.centrofermi.it/eee/).

The text is in Italian.

**The Author:** Costantino Sigismondi, astrophysicist and teacher, works on solar astrometry www.icra.it/solar.

**Cover image:** A simulated flux of muons coming from the zenith of Rome. The Gaussian field of view of the instrument has a FWHM (full width at half maximum) ~12°. The upper image is in equatorial coordinates and the lower in galactic coordinates. The 500 points distribution has been obtained with EEEtest.xls spreadsheet described in the text and downloadable here, and here. The images of our Galaxy are of A. Mellinger from the Calgary University website.

---

[1] M. Abbrescia, et al., *First detection of extensive air shower with the EEE experiment*, Nuovo Cimento B **125**, 243, (2010).

**Immagine di copertina:** Flusso di muoni provenienti dalla regione prossima allo zenith simulato per la latitudine di Roma. Il campo di vista gaussiano ha un'apertura FWHM di circa 12°. L'immagine superiore è in coordinate equatoriali, mentre quella inferiore in coordinate galattiche. La distribuzione di 500 punti è ottenuta col simulatore EEEtest.xls descritto nel testo e scaricabile [qui](), e [qui](). Le immagini della nostra Galassia di A. Mellinger sono dal sito dell'Università di Calgary (Canada).

**Prefazione:**

Avendo a che fare con i muoni, particelle che non possono essere riflesse né bloccate, poiché possono attraversare anche i 1500 metri di roccia sopra ai Laboratori Nazionali del Gran Sasso, ogni sistema di focalizzazione, collimazione o coded mask è inutile. Si può fare imaging dei raggi cosmici, ed individuare quindi le loro sorgenti celesti, solo ricostruendo la loro traiettoria di arrivo attraverso le tracce lasciate nei rivelatori multipiano MRPC (cfr. http://www.centrofermi.it/eee/ ).

I telescopi per muoni dell'esperimento EEE, Extreme Energy Events, montati nelle scuole per avvicinare gli studenti alla fisica dei raggi cosmici, hanno iniziato a registrare i primi eventi in coincidenza su due siti proprio nel 2008. Non c'è da meravigliarsi, dunque, se un algoritmo per passare dai coseni direttori alle coordinate altazimutali, equatoriali o galattiche non sia ancora disponibile né sui testi né sulla rete. Questo manuale vuole incominciare a colmare questo spazio ancora vuoto, ed il simulatore che lo accompagna consente di lavorare su dati artificiali anche coloro che ancora non dispongono dei rivelatori, avvicinandoli alle problematiche dell'astronomia di posizione di raggi cosmici ed anche dei neutrini.

**L'autore:** Costantino Sigismondi, astrofisico e docente, si occupa di astrometria solare alla Sapienza e all'Università di Nizza. Collabora dal 2008 con il Centro Fermi nell'ambito del progetto EEE, in particolare con il metodo di orientamento dei telescopi entro un minuto d'arco di precisione mediante l'uso delle Effemeridi solari, ha lavorato al simulatore del Satellite per Astronomia X italiano BeppoSAX ed al programma sul satellite francese PICARD per osservazioni solari.

Ateneo Pontificio Regina Apostolorum
Scienza e Fede GeoAstroLab
Manuali e Didattica
2

Costantino Sigismondi

# Astronomia di posizione per muoni

Algoritmi per foglio elettronico

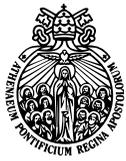
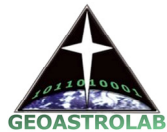

Figura in copertina: Flusso di muoni provenienti dalla regione prossima allo zenit simulato per le latitudini di Roma e l'Aquila.
Il campo di vista gaussiano ha un'apertura FWHM di circa 12°. L'immagine superiore è in coordinate equatoriali, mentre quella inferiore in coordinate galattiche. La distribuzione di 500 punti è ottenuta col simulatore EEEtest.xls descritto nel testo.
Le immagini della Galassia sono di Axel Mellinger (2000).

*Prima edizione: 23 Novembre 2008, Solennità di Cristo Re dell'Universo*

# Determinazione delle coordinate celesti di provenienza dei muoni cosmici a partire dai loro coseni direttori e tempo di arrivo: algoritmi per foglio elettronico

## Abstract

Positional Astronomy for Muons, Algorithms for Worksheet

A booklet seized for EEE, Extreme Energy Events
- http://www.centrofermi.it/eee/ - data format. From vectorial components of the incoming particles and time of the events we compute their celestial (equatorial and galactic) coordinates.

A worksheet file EEEtest.xls contains a simulator, to produce data in the same format.

This introductory method to positional astronomy for muons, and also neutrinos, is through explained formulae and an interactive worksheet.
**Author**
Costantino Sigismondi, astrophysicist and teacher, works on solar astrometry www.icra.it/solar .

## Introduzione

Avendo a che fare con i muoni, particelle che non possono essere riflesse né bloccate, poiché possono attraversare anche i 1500 metri di roccia del Gran Sasso, ogni sistema di focalizzazione, collimazione o *coded mask* è inutile. Si può fare *imaging* dei raggi cosmici, ed individuare quindi le loro sorgenti celesti, solo ricostruendo la loro traiettoria di arrivo attraverso le tracce lasciate nei rivelatori multipiano MRPC (cfr. http://www.centrofermi.it/eee/ ).

Solo con le "prime luci" dei telescopi per muoni in alcune scuole dell'Aquila e di Frascati nell'ambito del progetto EEE Extreme Energy Events promosso in Italia dal Centro Fermi, avvenute nel maggio del 2008 (Moro et al.), la realtà dell'astronomia dei raggi cosmici entra a contatto anche con gli studenti della scuola secondaria.

Non c'è da meravigliarsi, dunque, se un algoritmo per passare dai coseni direttori alle coordinate altazimutali, equatoriali o galattiche non sia ancora disponibile né sui testi né sulla rete.

Questo manuale vuole incominciare a colmare questo spazio ancora vuoto.

Il simulatore che lo accompagna consente di lavorare su dati artificiali anche coloro che ancora non dispongono dei rivelatori, permettendo pure di verificare in anticipo la loro resa in funzione della separazione tra i piani che determina il campo di vista dello strumento, e della porzione di piano galattico visibile.

Col simulatore ci si può avvicinare più facilmente alle problematiche dell'astronomia di posizione dei raggi cosmici ed anche dei neutrini, applicandone i principi anche alle osservazioni fatte nei grandi laboratori dedicati sia sotterranei come il Gran Sasso che sottomarini, sia al grande osservatorio Pierre Auger (Matthiae, 2008) nella pampa Argentina (si consulti Bernloehr, 2008 per una panomarica completa di tutti gli esperimenti su raggi cosmici, gamma e neutrini).

Inoltre l'astronomia sferica classica è sempre meno insegnata nelle università a vantaggio dell'astrofisica, ed in questo testo si ha invece l'occasione di riprendere temi base di astronomia sferica utilizzando i metodi dell'algebra lineare comuni a tutti i corsi di geometria del primo anno delle facoltà scientifiche.

Un evento osservato da telescopi per muoni cosmici, come quelli in uso nell'esperimento EEE, Extreme Energy Events, è caratterizzato dai suoi coseni direttori e dall'istante di tempo dell'osservazione. I coseni direttori individuano la direzione di provenienza dello sciame muonico nel sistema altazimutale (altezza sull'orizzonte locale e distanza angolare dal meridiano) mentre dal dato temporale si risale all'ascensione retta del meridiano celeste che in quel momento è a Sud (tempo siderale). Calcolando la distanza del punto osservato dal Polo Nord Celeste (PNC, fisso) si ottiene la sua declinazione come angolo complementare a 90°, mentre dalla sua distanza dal meridiano chiamata angolo orario (misurata su un parallelo celeste centrato nel PNC) si ottiene l'ascensione retta. Declinazione ed ascensione retta sono le coordinate equatoriali in astronomia, determinate dalla direzione assoluta dell'asse terrestre; una matrice di rotazione consente di passare alle coordinate galattiche, determinate dal piano della Via Lattea.

Si descrivono tutti i calcoli per arrivare ad impostare su un foglio di lavoro l'algoritmo per conoscere le sorgenti celesti di EEE. Questo approccio è pensato per consentire anche agli studenti di una scuola secondaria di comprendere tutti gli aspetti del problema.

## Allineamento dei telescopi con il Nord celeste

E' possibile orientare i telescopi per muoni posti nelle scuole con una precisione migliore di un minuto d'arco con piccoli accorgimenti.

Si può iniziare con l'orientare un muro esterno della scuola, che assumiamo parallelo (o perpendicolare) a quello interno della stanza dove è collocato il telescopio.

Occorre conoscere tramite GPS o Google maps le coordinate geografiche del punto, latitudine e longitudine, e l'istante di tempo in cui la luce del Sole è radente a questo muro esterno.

Come programma di effemeridi consiglio Ephemvga, disponibile con le istruzioni per l'uso al sito www.santamariadegliangeliroma.it menù meridiana – calcolo delle effemeridi; le istruzioni sono anche più dettagliate sul mio testo (2008) Effemeridi.

Ogni programma di Effemeridi fornisce ad ogni istante l'azimut del Sole, che è riferito al Polo Nord Celeste, dandoci una misura non affetta dalle anomalie magnetiche locali.

Dall'orientamento del muro a quello del telescopio si arriva lavorando su triangoli all'interno della stanza con i metodi della geometria piana per conoscere gli angoli di un triangolo di cui sono noti i 3 lati (teorema del coseno: $c^2=a^2+b^2-2ab\cdot\cos\gamma$).

Un altro metodo, simile, utilizza l'ombra proiettata all'interno della stanza da un bordo verticale di una finestra. Questa linea a terra è in continuo lento movimento, fissandone la direzione ad un istante ben preciso, con il programma di effemeridi si conosce l'orientamento assoluto di quella linea.

Poi si procede mediante la soluzione di triangoli di cui si conoscono 3 lati di cui il primo coincide con la direzione dell'ombra, il secondo con quella del telescopio, ed il terzo unisce le due direzioni.

# Tempo Siderale

Dato un istante di tempo in cui si osserva un EEE, dato in ora locale, oppure tempo universale coordinato (UTC), il tempo siderale locale TSL è la coordinata del meridiano celeste che in quel momento transita al meridiano locale.

Un giorno siderale dura 23 h 56 m 04.09 s, quanto la rotazione terrestre. Perciò occorre conoscere un valore del tempo siderale per un dato luogo e ad un dato istante UTC per poter calcolare, sempre per quel luogo ad ogni altro istante il TSL. L'Unione Astronomica Internazionale ha così definito l'origine del TS di Greenwich

$TS_{0\ Greenwich}$ (alle 0 UTC di ogni giorno)= 6 h 39 m 52.26 s + 8640184.812866·T [s]

con T misurato in secoli giuliani (36525 giorni) a partire dalle 0 UTC del 1 gennaio 2000.

La nutazione dell'asse terrestre induce una piccola oscillazione chiamata Equazione degli Equinozi (EE) dell'ordine di 1 s, che dà il TS vero; rinvio ad una trattazione più completa (Barbieri, 1999 cap. 10) per approfondimenti di astrometria di precisione.

Dato che la precisione angolare che possiamo ottenere sulla direzione di provenienza dei µ non è migliore di 1′, l'equazione degli equinozi può essere trascurata. Infatti vale la proporzione 1 ora di ascensione retta = 15° angolari, 1 minuto di a. r. = 15′, 1 secondo di a. r. = 15″ angolari. Dunque 1′ angolare corrisponde a 4 s di tempo.

## Esempio di calcolo per TS

Consideriamo le 11:00 UTC dell' 11 novembre 2008. A Roma 41°54′ N, 12°30′ E.

Conviene usare la data giuliana (JD), che si può trovare con ephemvga o sul web.

JD 11/11/2008 h. 00:00 UTC è 2454781.5

JD 01/01/2000 h. 00:00 UTC è 2451544.5

La differenza è 3237 giorni.

Si calcola il T=3237/36525=0.088624229 in secoli giuliani trascorsi dal 1/1/2000.

Si calcola il valore $TS_0$ Greenwich (11/11/2008 0 UTC)= 6 h 39 m 52.26 s +765729.73 s, il termine lineare in T si divide per 86400 s in un giorno e si prende solo la parte eccedente l'intero, trasformandola in ore minuti e secondi. Risultano 20 h 42 m 05.73 s, che vanno sommati al valore di zero, ed il risultato viene preso modulo 24 ore. Risulta $TS_0$ Greenwich (11/11/2008 0 UTC)=3 h 22 m 01.99 s.

Poiché un giorno siderale vale 86164.09 s di tempo solare medio, 11 ore di tempo solare medio corrispondono ad 11·(86400/86164.09) ore di TS = 11 h 1 m 48.42 s che sommate ad TS (0 UTC dell'11/11/2008) danno TSL Greenwich (11 UTC 11/11/2008) = 14 h 23 m 50.41 s.

Infine 12°30′ corrispondono a 50 minuti di tempo siderale, per cui a Roma in quel momento il TS vale

TSL Roma (11:00 UTC 11/11/2008)= 15 h 13 m 50.41 s.

## Declinazione

I dati ottenuti con i telescopi per muoni del progetto EEE sono già sotto forma di coseni direttori. Di ogni direzione di provenienza vengono fornite le componenti vettoriali di norma unitaria, cioè le proiezioni del vettore a, b, c sugli assi coordinati X (diretto verso Nord) Y (diretto verso Est) e Z (diretto verso lo zenit) tali che $a^2+b^2+c^2=1$.

Di solito in astronomia nautica e sferica questi dati vengono forniti in altezza (distanza in gradi dall'orizzonte) ed azimut (distanza in gradi dal punto cardinale Nord letta in senso orario da Nord verso Est), poiché venivano misurate tramite i sestanti progettati proprio per dare queste misure.

Con l'avvento dell'astronomia dei raggi cosmici per la prima volta le particelle che portano l'informazione della direzione del raggio cosmico primario non vengono rivelate solo nella loro posizione di arrivo, ma la loro direzione è campionata attraverso una traccia che lasciano nei telescopi ad essi dedicati.

Questo processo di abbandono del concetto tradizione di "imaging" ottenuto mediante la focalizzazione con lenti o specchi dei fotoni provenienti da sorgenti poste all'infinito, ha avuto inizio con l'astronomia X dove l'uso di collimatori e coded masks consentiva di identificare la regione di cielo di provenienza entro qualche grado di precisione.

Specchi ad incidenza radente (Sigismondi, 1997) hanno consentito ai satelliti per osservazioni nei raggi X di fare imaging con risoluzioni spaziali del secondo d'arco (XMM Newton), tuttavia all'aumentare dell'energia dei fotoni gli angoli di incidenza per cui il fotone viene riflesso dalle superfici dorate tende a zero, e solo sistemi concettualmente analoghi ai collimatori vengono usati per i satelliti dedicati alle osservazioni gamma (Tavani et al, 2004).

Quando le particelle in oggetto sono i muoni, essi possono attraversare grandi quantità di materia senza essere assorbiti, e solo la loro traccia può essere identificata mediante opportuni rivelatori.

In questo caso l'imaging non è fatto mediante focalizzazione di un fascio di particelle collimate all'infinito, giacché essi non possono essere riflessi da nessuna superficie riuscendo ad attraversare anche una montagna come il Gran Sasso. I raggi cosmici secondari non sono neppure collimati all'infinito, poiché sono prodotti all'ingresso dell'atmosfera terrestre.

Non si può fare *imaging* neppure tramite apparati collimatori o *coded masks* poiché, per le stesse ragioni precedentemente esposte attraversano ogni tipo di *mask*.

Avendo, dunque, i coseni direttori di provenienza, si può calcolare mediante semplici prodotti scalari gli angoli di arrivo rispetto al meridiano, detto angolo orario, e rispetto al polo nord celeste, che è il complementare della declinazione.

Il calcolo del prodotto scalare con il foglio elettronico è assai semplice e praticamente scevro da errori anche tipografici.

Per la declinazione occorre identificare i coseni direttori del Polo Nord Celeste (PNC), che è sempre fisso anche nel sistema di riferimento locale orizzontale.

Il PNC è un vettore di raggio 1 che nella sfera celeste parte dal centro e punta a Nord formando un angolo con l'orizzonte pari alla latitudine $\varphi$ del luogo.

Le sue componenti sono    PNC $(\cos\varphi, 0, \sin\varphi)$.

Il prodotto scalare è dunque $w = a\cdot\cos\varphi + c\cdot\sin\varphi$

Se sulla colonna A ci sono i valori a del primo coseno direttore e nella C quelli del terzo coseno direttore di ogni vettore, $D1 = w1 = A1*\cos\varphi + C1*\sin\varphi$, dove $\cos\varphi$ e $\sin\varphi$ sono numeri.

Questo numero $w1$ è il coseno dell'angolo compreso tra le due direzioni dello spazio.

La declinazione $\delta = 90° - \arccos(w1)$

$E1 = 90 - \text{gradi}(\arccos(D1))$

Dove la funzione "gradi" trasforma in gradi il risultato, dato in radianti, uscente dalla funzione "arccos".

# Ascensione Retta

Altrettanto semplice il caso dell'angolo orario per arrivare all'ascensione retta.

L'angolo formato tra la direzione di provenienza ed il meridiano può essere ottenuto immediatamente dal prodotto scalare della proiezione del vettore sul piano orizzontale con il punto cardinale Sud (-1,0,0).

Quindi il prodotto scalare w2=-a, e l'angolo è h=F1=gradi(arccos(-A1))

Il valore di w2 è compreso tra -1 ed 1, quando l'angolo orario è compreso tra 180° (direzione esattamente a Nord) e 0° (esattamente a Sud), tuttavia a valori intermedi come, ad esempio, con a=-0.5, w2=0.5, ah=60°, ma può essere sia a Sud-Est che a Sud Ovest. La discriminazione tra i due casi avviene considerando che l'azimut a=180°-ah se b>0 oppure a =180°+ah se b<0.

Nel foglio elettronico questa condizione si realizza con una funzione "if", che matematicamente corrisponde alla moltiplicazione per il segno di b:

=GRADI(ARCCOS(-H2/RADQ(H2^2+I2^2)))*SEGNO(I2)/360

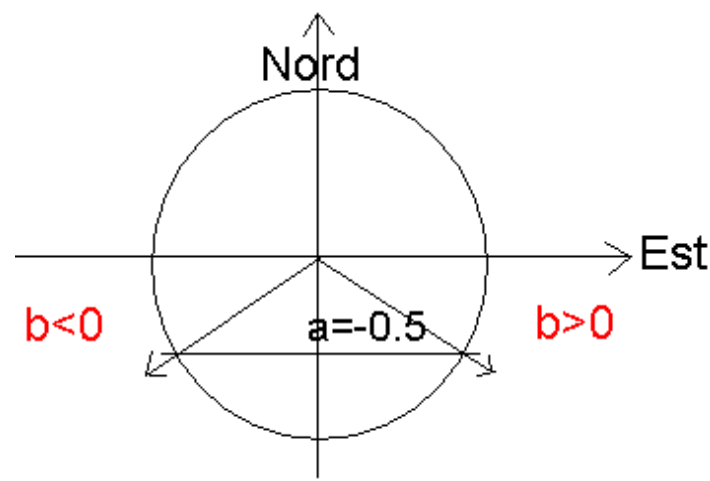

Per risalire all'ascensione retta a partire dall'angolo orario, letto col segno positivo in senso antiorario a partire dal Sud, si somma al tempo siderale all'istante dell'osservazione l'angolo orario; del risultato finale si prende solo la parte eventualmente eccedente le 24 ore.

Per comodità di calcolo si possono svolgere i calcoli in ore, oppure gradi dell'angolo giro, oppure frazioni decimali di angolo giro.

=360*((K2-L2)-INT(K2-L2))

Ad esempio nel caso a=-0.5 e b>0, con TSL=1 h abbiamo ah=3h, azimut a=120° e l'ascensione retta della sorgente è $\alpha$= 4 h, perché passerà al meridiano 3 h dopo.

Nel caso di b<0 e stesso valore di a=-0.5, la sorgente era passata già 3 h prima al meridiano, quindi ha un'ascensione retta pari ad $\alpha$=22 h. In questo caso ah=-60°, e l'azimut a=240°.

# Passaggio alle coordinate galattiche

Il passaggio da un sistema di riferimento celeste all'altro non è altro che una rotazione in cui, ad esempio, il Polo Nord Celeste viene sostituito da quello Galattico.

Nel nostro caso questo passaggio serve a visualizzare eventuali legami tra le coordinate di arrivo dei muoni ed il piano galattico. Le coordinate galattiche stanno a quelle equatoriali come in figura:

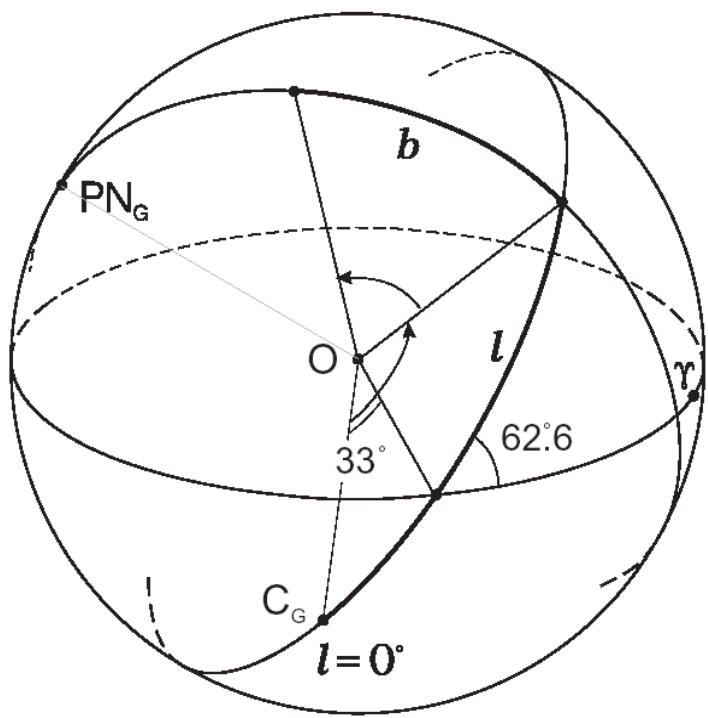

b è la latitudine galattica, l è la longitudine contata dalla direzione del Centro Galattico in senso antiorario; in queste coordinate il punto γ, origine delle coordinate equatoriali, possiede l=33° (al 1950.0).

Nel sistema di coordinate galattiche l'asse X è rivolto verso il centro galattico (qui tratteggiato fino) e Z verso il Polo Nord Galattico (linea sottile continua). L'inclinazione del piano galattico sull'equatore celeste è pari ad i=62.6°, non definito in figura.

Le coordinate di Polo e Centro Galattico sono all'epoca 2009.0

PNG (12 h 52.7 m, 27°06′)

CG (17 h 47.0 m, -28°56′)

Poiché il piano dell'equatore galattico non è ben definito, esso viene assunto come convenzione, perciò le coordinate galattiche non vengono mai usate come misure di posizione precisa.

Per gli scopi di questo testo è sufficiente calcolare la latitudine galattica a partire dalle coordinate equatoriali usando la formula di trigonometria sferica (aggiornata al 2009.0) facilmente implementabile sul foglio elettronico:

senb=sen(27°06′)·sen(δ)+cos(27°06′)· cos(δ)·cos(α-193°10′30″)

b=GRADI(ARCSEN(0.455544507*SEN(RADIANTI(M4))+0.890 2126*COS(RADIANTI(M4))*COS(RADIANTI(N4-193.175))))

mentre per ricavare anche la longitudine galattica occorre ricavare dalle altre due relazioni (cfr. Barbieri, 1999) della trigonometria sferica la seguente relazione (Duffet-Smith, 1983 aggiornata al 2009.0):

$$l = \arctan\left(\frac{\text{sen}(\delta) - \text{sen}(b)\cdot\text{sen}(27°06)}{\cos(\delta)\cdot\text{sen}(\alpha - 193°1030") \cdot \cos(27°06)}\right)$$

Per risolvere l'ambiguità dell'arcotangente (y/x) occorre determinare i segni di y ed x con la funzione SEGNO, poi aggiungere 0°, 180° o 360° a seconda dei casi, come nella figura seguente.

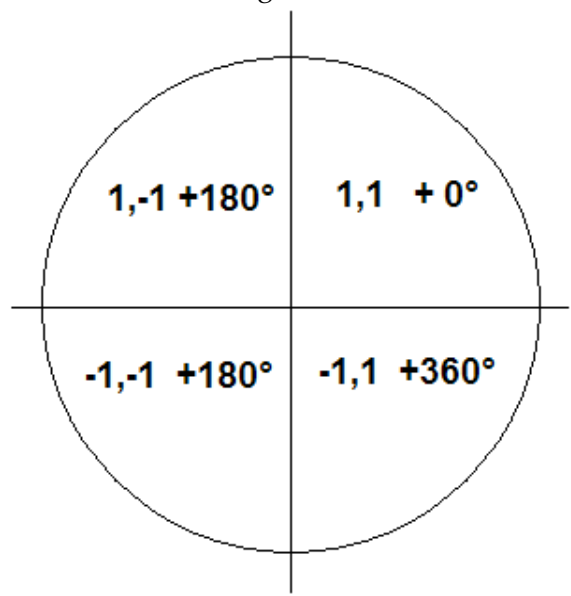

Nel foglio elettronico questo termine additivo si può ottenere con la seguente somma di formule logiche (nelle caselle r3 ed s3 sono i segni di y ed x rispettivamente).

=SE(E(R3=1;S3=1)=VERO;0;FALSO)+SE(E(R3=-1;S3=-1)
=VERO;180;FALSO) +SE(E(R3=-1;S3=1)=VERO;360;FALSO)
+SE(E(R3=1;S3=-1)=VERO;180;FALSO)

## Simulatore

Un simulatore consente di riprodurre i dati osservati mediante un software, che, quanto meglio tiene conto di tutte le caratteristiche dello strumento e del fenomeno in esame, tanto meglio può anche consentire di mettere a punto miglioramenti, programmi di analisi dati e nuove strategie osservative.

Il file EEEtest.xls ha in input il tempo, in secondi dall'inizio dell'anno 2008 (incluso il leap second alla fine dell'anno 2008 (IERS, 2008),

A3=31622401*CASUALE() ,

ed i coseni direttori scelti attorno allo zenit, o attorno ad un'altra direzione nello spazio.

Nel caso di finestra gaussiana attorno allo zenit abbiamo

D3=1-ASS(INV.NORM(CASUALE();0;0.1))

la varianza pari a 0.1 corrisponde ad una larghezza FWHM, Full Width at Half Maximum, di circa 30°, simile a quella osservata nelle scuole dell'Aquila (Moro et al., 2008).

Con quest'altra funzione

D3=0.739+(INV.NORM(CASUALE();0;0.004))

si descrive la linea dell'equatore celeste (alla latitudine di L'Aquila e Roma), se gli altri due coseni direttori sono fissi a -1 e 0.

La proiezione sul piano orizzontale e la norma unitaria del vettore si realizza con le funzioni

B3=RADQ(1-D3^2)*COS(RADIANTI(A3))

C3=RADQ(1-D3^2)*SEN(RADIANTI(A3))

dove in A3 c'è un numero casuale, comune per le due componenti B3 e C3.

Questi dati in input sono nello stesso formato di quelli reali dell'esperimento EEE.

I risultati di queste simulazioni sono rappresentati nella figura di copertina.

# Didattica

Per convincersi della bontà dell'analisi dei dati, in particolare della distribuzione ad anello che appare nelle coordinate galattiche, con relativa scarsità di eventi tra 180° e 360° di longitudine galattica, basta usare un comune globo mappamondo.

L'equatore rappresenta l'equatore celeste. Un nastro è fissato in modo da descrivere un cerchio massimo che interseca l'equatore con un angolo di 62°, questo nastro rappresenta l'equatore galattico, il polo del mappamondo è il Polo Nord Celeste.

Inclinando a circa 45° l'asse terrestre e guardando dall'alto vediamo le coordinate della regione di cielo corrispondente allo zenit.

Ebbene con la rotazione diurna allo zenit viene spazzato una regione attorno ad un parallelo celeste (e quindi anche terrestre).

Il parallelo celeste a 42° dal Nord, 48° dall'equatore (per Roma e l'Aquila), se è sottile, intercetta l'equatore galattico per un breve tratto, e dall'altra parte non raggiunge il Polo Galattico, escludendo quindi una vasta regione di longitudini galattiche.

La foto seguente chiarisce la semplicità di questo apparato didattico.

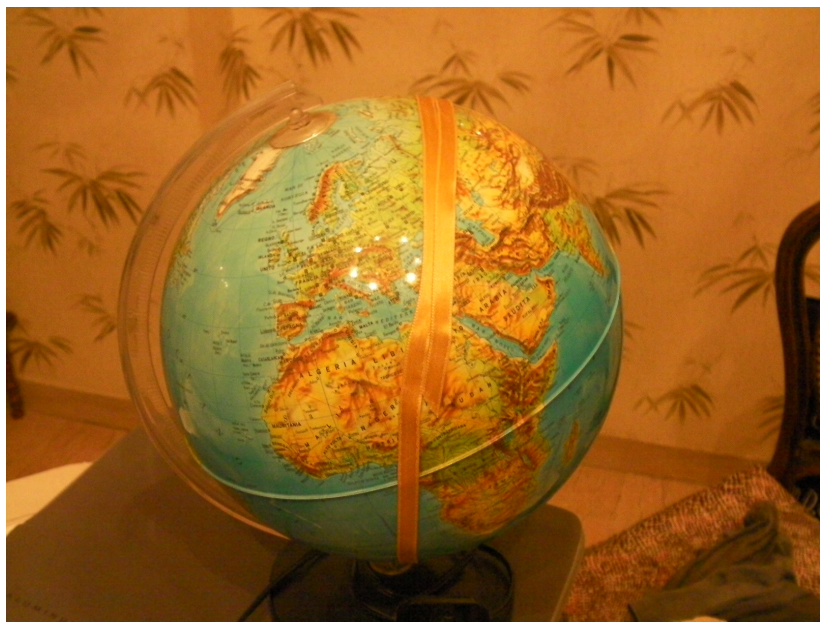

# Ringraziamenti

A Rinaldo Baldini Ferroli, Direttore del Centro Fermi, che ha subito apprezzato il metodo di allineamento dei telescopi con le effemeridi solari e mi ha stimolato a studiare questi algoritmi ed a Roberto Moro, dei Laboratori Nazionali del Gran Sasso, che mi ha dato in anteprima i dati originali delle scuole Aquilane: Liceo Scientifico Andrea Bafile, e ITIS Amedeo di Savoia Duca d'Aosta.

A Giorgio Sordello, già Preside dell'IISS Alessandro Volta di Roma, grazie al quale sono entrato in contatto con l'esperimento EEE.

# Referenze

Juergen Giesen (2007)
http://www.jgiesen.de/SiderealTimeClock/index.html è un Applet Java per il calcolo del TS per ogni luogo con precisione al centesimo di grado di longitudine.

US Naval Observatory Time Service (2008)
http://tycho.usno.navy.mil/sidereal.html permette il calcolo del TS apparente istantaneo (include l'equazione degli equinozi).

Elwood Charles Downey (1992) Ephemvga 4.27, software free su www.santamariadegliangeliroma.it menù meridiana, calcolo delle effemeridi, calcola il TS con precisione al secondo d'arco di longitudine, consente anche il calcolo della precessione alle coordinate equatoriali.

Rajiv Gupta (2000), Observer's Handbook 2001, University of Toronto Press (Toronto, Canada), p. 35, fornisce al decimo di secondo il TS medio di Greenwich al giorno 0, alle ore 0 UTC di ogni mese.

# Indice